\documentclass[aip,adv,reprint,twocolumn,nofootinbib]{revtex4-2}
\usepackage{amsfonts,amsmath,amsthm,enumitem}
\usepackage{xspace}
\usepackage{graphicx}

\newcommand{\AO}{AO\xspace}
\newcommand{\QKD}{QKD\xspace}
\newcommand{\QBER}{QBER\xspace}
\newcommand{\LEO}{LEO\xspace}
\newcommand{\TDM}{TDM\xspace}
\newcommand{\DM}{DM\xspace}
\newcommand{\WDM}{WDM\xspace}
\newcommand{\FS}{FS\xspace}
\newcommand{\BB}{BB84\xspace}

\begin{document}
\title{Reducing The Impact Of Adaptive Optics Lag On Optical And Quantum Communications Rates From Rapidly Moving Sources}
 
\author{Kai Sum Chan}
\altaffiliation{Present address: Quantum Bridge Technologies Inc., 100 College
 Street, Toronto, Ontario, Canada and National Research Council Of Canada,
 Ottawa, Ontario, Canada}
\author{H.~F. Chau}
\email[Corresponding author, email: ]{{\tt hfchau@hku.hk}}
\affiliation{Department of Physics, University of Hong Kong, Pokfulam Road,
 Hong Kong}
\date{\today}

\begin{abstract}
 Wavefront of light passing through turbulent atmosphere gets distorted.
 This causes signal loss in free-space optical communication as the light beam spreads and wanders at the receiving end.
 Frequency and/or time division multiplexing adaptive optics (\AO) techniques have been used to conjugate this kind of wavefront distortion.
 However, if the signal beam moves relative to the atmosphere, the \AO system performance degrades due to high temporal anisoplanatism.
 Here we solve this problem by adding a pioneer beacon that is spatially separated from the signal beam with time delay between spatially separated pulses.
 More importantly, our protocol works irrespective of the signal beam intensity and hence is also applicable to secret quantum communication.
 In particular, using semi-empirical atmospheric turbulence calculation, we show that for low earth orbit satellite-to-ground decoy state quantum key distribution with the satellite at zenith angle $< 30^\circ$, our method increases the key rate by at least $215\%$ and $40\%$ for satellite altitude $400$~km and $800$~km, respectively.
 Finally, we propose a modification of existing wavelength division multiplexing systems as an effective alternative solution to this problem.
\end{abstract}
\maketitle

\section{Introduction}
\label{Sec:Intro}

 Optical free-space communications and astronomical imaging are affected by
 atmospheric turbulence due to fluctuation of air density, pressure, and
 temperature. 
 This turbulence induces a time-dependent inhomogeneous refractive index in
 air, distorting the wavefront of electromagnetic waves.
 Hence, light beam spreads and wanders at the detection end causing signal
 loss.
 High fidelity signal or image is obtained if one could adaptively and
 dynamically conjugate the optical path difference caused by the wavefront
 distortion.
 Adaptive optics (\AO) is a well established method to achieve this
 goal~\cite{Roddier,Guyon,Tyson}.
 In the most basic \AO setup, a deformable mirror (\DM) is used to collect light
 signal and a beam splitter is placed in front of the signal detector acting
 as a signal sampler to divert some signal light to a wavefront sensor.
 The detection result of this sensor is then used to estimate the wavefront
 distortion.  Finally, one can adaptively conjugate this estimated distortion
 via fast feedback control of the \DM through actuators to obtain a high fidelity signal or
 image~\cite{Roddier,Guyon,Tyson,WangAO18}.

 Many variants of this basic setup have been proposed and used in the field.
 For instance, one may replace the beam splitter by a wavelength selector plus
 an additional beacon beam emitting light with a different wavelength from
 that of the signal beam.  This wavelength-division multiplexing (\WDM) setup
 is effective if the wavelengths of the two beams are close enough so that the
 wavefront distortion inferred from the beacon beam is close to that of the
 signal beam.  And at the same time, the wavelength difference is big enough
 to avoid cross-talk between the two beams.
 Another variant is to use time-division multiplexing (\TDM) method in which
 the beam splitter is replaced by an optical switch and a pulsed beacon
 beam~\cite{Tyson}.  We remark that in most \WDM and \TDM setups, the beacon
 and the signal beams are spatially coincided.
 In order to work, the \WDM and \TDM methods must use a sufficiently high
 intensity beacon beam so that the wavefront sensor can detect enough photons
 per unit time to estimate the wavefront distortion accurately.  In contrast,
 the brightness of the signal source is irrelevant as far as \AO correction is
 concerned.  That is to say, both \WDM and \TDM methods work
 for low intensity signal sources, including most quantum signal sources used
 in secure quantum communication.  In fact, \WDM has been used in a few
 recent quantum communication experiments~\cite{Cao20,Gruneisen_2021}.
 
 A new challenge is faced if the signal source moves sufficiently fast relative to the atmosphere.
 This increases the temporal angular distance between the optical path of the beacon and the corresponding path of the signal due to \AO lag.
 The temporal anisoplanatism induced by the movement of the source greatly degrades the system's performance.
 Our method to tackle this challenge is inspired by astronomical imaging of
 dim celestial objects.
 Recall that astronomers use an artificial high intensity laser guide star
 placed angularly close to the dim astronomical object as beacon source to
 replace the role of the diverted signal light~\cite{Roddier,Guyon,KeckAO}.
 With this inspiration, we solve the moving source problem by using two set of
 spatially separated artificial sources emitting at the same or nearly the same
 wavelength --- a set of (pulsed) pioneer beacon source(s) to perform
 effective \AO correction and another set of time-delayed (pulsed) signal
 source(s) for the actual optical communication.
 In essence, our proposal is a time-delayed spatial multiplexing protocol.
 This protocol can also be interpreted as a time-based \AO pre-compensation scheme in the sense that the \DM pre-deforms before the signal arrives.
 In this different from another type of \AO pre-compensation scheme in uplink satellite communication in which the signal is pre-shaped before sending to the satellite~\cite{uplink1,uplink2}.

\begin{figure}[t]
\centering
\includegraphics[width=0.4\textwidth]{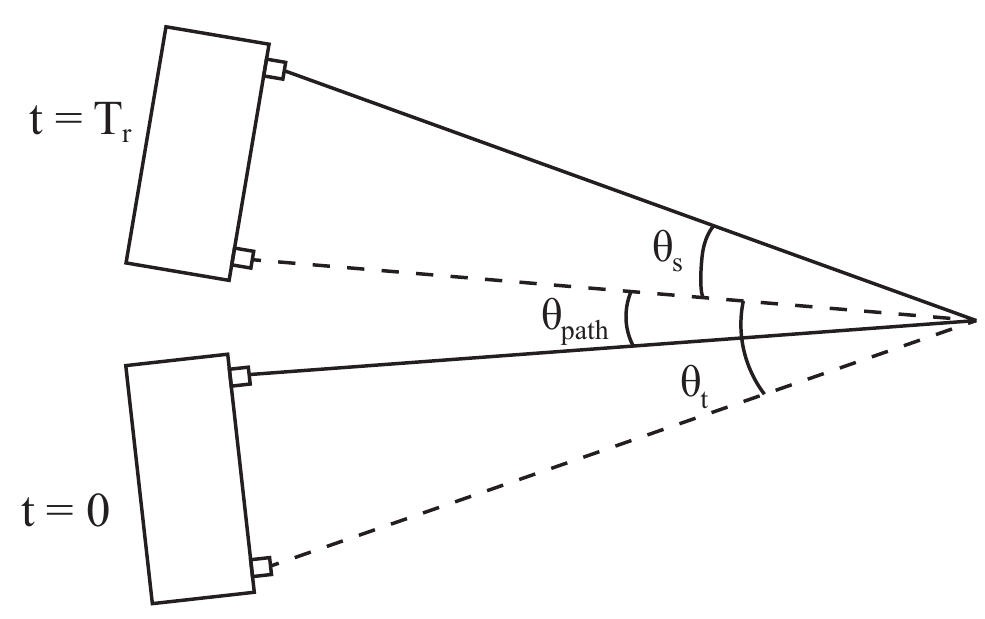}
 \caption{Satellite location and beams' path at $t = 0$ and $t = T_r$.
 Here $\theta_s$ is the angular separation of the pioneer beacon and signal beams (as observed from the receiver), and
 $\theta_t$ is the angle traveled by the signal (pioneer beacon) beam's path
 from $t = 0$ to $t = T_r$ [see the solid (dash) lines].
 In addition, $\theta_\text{path}$ is the angle between the pioneer beacon 
 beam's path (solid line at $t = 0$) and the delayed signal beam's path (dash line at $t = T_r$).}
\label{Fig:time}
\end{figure}

 For concrete illustration, we consider the following prototype from now on
 although the general concept works in a much wider context.
 As shown in Fig.~\ref{Fig:time},
 we consider the satellite-to-ground communication setup with both the beacon and
 signal sources are fixed on a low earth orbit (\LEO) satellite together with a
 stationary ground-based receiver telescope.
 By pioneer beacon source, we mean that the beacon beam is put in front of the
 signal beam along the direction of motion of the satellite relative to the
 ground.
 Furthermore, we fire each pulsed pioneer beacon beam shortly before firing
 the corresponding pulsed signal beam.
 In doing so, the beacon beam acts like a pioneer that probes the wavefront
 distortion of an optical path that will shortly be traveled by the signal
 beam.
 Specifically, if the two beams move sufficiently rapidly relative to the
 detector(s), the pulsed pioneer beacon beam and the corresponding delayed
 pulsed signal beam can be made to travel along essentially the same optical
 path by carefully tuning the delay time.
 Consequently, our time-based \AO pre-compensation technique should achieve almost the same level of \AO correction for
 stationary sources without spatial multiplexing.
 As the two light sources are multiplexed spatially, their signals can be separated
 effectively by focusing the light beams provided that the angular separation
 of their images after applying \AO correction is greater than the resolving
 power of the ground-based telescope and that the cross-talk between them is sufficiently small.
 In fact, a recent experiment using the 1~m telescope in Mount Stromlo
 Observatory plus \AO imaging technique succeeded to image an artificial
 satellite down to 85~cm in size at 1000~km range~\cite{Bennet16}.  This
 implies that using telescope with aperture larger than about 1~m, our
 prototype is able to resolve and separate the pioneer beacon and signal beams
 mounted on a typical-sized artificial satellite.

 Note that we study this prototype because this is one of the most challenging
 situations in realistic applications.  As the effectiveness of our
 approach does not depend on the nature of the signal light source, so by the
 same logic, we choose our signal source to be a phase-randomized weak
 coherent quantum source
 performing decoy state quantum key distribution (\QKD).  In this way, we
 could demonstrate the strength of our approach and compare it with existing
 ones.
 In fact, free-space channel is used in quantum communication because it has
 a lower attenuation rate than optical fibers of the same
 length~\cite{Pirandola_2020}.
 No wonder why several pioneering demonstrations of long distance free-space
 \QKD, including ground-to-ground and satellite-to-ground ones, have been
 reported~\cite{Liao_2017,Xu_2020,Cao20,Gruneisen_2021}.
 For free-space \QKD, existing \AO technologies are able to increase the key
 rate by reducing the widening effect and spatial noise of the signal so that
 the system can get a higher yield or coupling efficiency even in
 daytime~\cite{Gruneisen_2021,Gruneisen2014}.
 And to the best of our knowledge, all \AO-based free-space \QKD experiments
 to date use \WDM~\cite{Cao20,Gruneisen_2021}.
 A drawback of this approach is that the different wavefront distortions
 experienced by the beacon and signal beams generally increases with
 communication distance.  This could lower the yield and key rate when this
 distance is long.  
 More importantly, both \WDM and \TDM suffer from huge temporal anisoplanatism.

 We begin by presenting the atmospheric model and system parameters used in our investigation in Sec.~\ref{Sec:Model}.
 Then in Sec.~\ref{Sec:Time}, we introduce our time-delayed spatial multiplexing method of time-based \AO pre-compensation that uses a pulsed pioneer beacon beam plus a time-delayed pulse signal beam.
 We also analyze its effectiveness in transmitting information through the dynamical atmosphere.
 In Sec.~\ref{Sec:Spatial}, we show the schematic design of the spatial multiplexing system and discuss the cross-talk due to the pioneer beacon.
 With the above preparatory works, we study the performance of our scheme
 for the case when cross-talk between the pioneer beacon and the signal beams
 can be ignored in Sec~\ref{Sec:compare}.
 Specifically, for the case of our concrete illustrative example, our scheme
 always gives higher Strehl ratio and transmission efficiency over those that
 use pure \TDM or \WDM.
 Besides, we analyze situation in which cross-talk cannot be ignored in Sec.~\ref{Sec:QKD}.
 There we compute the secret key rate of decoy state \BB \QKD that is optimized over signal beam parameters.
 Again, we find that for our concrete illustration, our scheme always gives a higher provably secure key rate over pure \TDM and \WDM protocols.
 By semi-empirical calculation, we find that for satellite at zenith angle $< 30^\circ$, the provably secure \QKD key rate of our scheme is increased by at least $215\%$ and $40\%$ when the satellite altitude is $400$~km and $800$~km, respectively.
 We also find that generally a greater key rate improvement is obtained when the system bandwidth is lower, the distance between the pioneer beacon and signal beams is higher, and the angular speed of the satellite relative to the ground detector is faster.
 As our setup is new and its construction is engineering demanding, a compromise is to upgrade existing \WDM systems to combine the beacon and signal beams.
 We report the performance of this modification in Sec.~\ref{Sec:WDM_improve}.
 We find that for zenith angle less than about $45^\circ$, the provably secure key rate of this modification is at least $90\%$ of our spatial multiplexing
 prototype reported in Sec.~\ref{Sec:QKD}, making it an attractive practical alternative.
 Finally, we summarize our findings in Sec.~\ref{Sec:conclusion}.
 The present work is based on our recent
 patent application~\cite{patent} and the Master thesis of one of the
 authors~\cite{chan_thesis}.
 
\section{Atmospheric Model and System Parameters of the Free-Space Communication Channel}
\label{Sec:Model}
\subsection{Atmospheric Model}
\label{Subsec:Atmosphere}
The Fried parameter $r_0$ is one of the most important quantity characterizing the atmospheric coherence diameter due to turbulence~\cite{Fried:65}.
In unit of meters, its value varies with the altitude and the zenith angle according to the equation
\begin{equation}
\label{E:fried}
 r_0 = \left [0.423k^2 \sec(\zeta )\int_{0}^{h_\text{alt}} C^2_n(h)\,dh\right ]^{-3/5},
\end{equation}
where $\zeta$ is zenith angle, $k$ is the wavenumber 
of the light measured in m$^{-1}$, $h_\text{alt}$ is the altitude of the source measured in meters, and
$C^2_n(h)$ is the refractive index structure parameter at altitude $h$.
Here we assume that
$C^2_n(h)$ follows the Hufnagel-Valley model~\cite{Hufnagel:64}, namely,
\begin{align}
 & C^2_n(h) \nonumber \\
 ={} & 0.00594\left(\frac{w}{27}\right)^2\left(\frac{h}{10^5}\right)^{10}\exp\left(-\frac{h}{1000}\right) \nonumber \\
 & +2.7\times 10^{-16}\exp\left(\frac{-h}{1500}\right)+ 1.7\times 10^{-14}\exp\left(\frac{-h}{100}\right)
 \label{E:C2_def}
\end{align}
with $h$ measured in meters.
In most literature, $w = 21$~m/s is the pseudo-wind speed, taken to be the average wind speed of the jet stream~\cite{Hufnagel:64}.
We stress that Eqs.~\eqref{E:fried} and~\eqref{E:C2_def} are valid for a source that is either stationary or moving relative to the detector.

Isoplanatic angle $\theta_0$ and Greenwood frequency $f_G$ are two quantities that can be used to characterize the spatial and temporal limits in \AO.
At the receiving end, light rays coming from a cone with an angle much smaller than $\theta_0$ has about the same optical path length.
And $f_G$, which is the reciprocal of the beam wandering time, is an effective way to approximately quantify the rate of change of turbulence~\cite{Greenwood:77,Tyson}.
Clearly, for a stationary source, \AO is effective only if the angular separation between the (pulsed) beacon beam and the (pulsed) signal beam is much less than $\theta_0$.  Moreover, the time delay between these two pulsed beams is much less than $1/f_G$.
These two quantities can be computed via $C^2_n(h)$ through
\begin{equation}
\label{E:theta_0}
    \theta_0 = \left[ 2.913k^2\sec^{8/3}(\zeta) \int_{0}^{h_\text{alt}}h^{5/3}C^2_n(h)\,dh \right]^{-3/5}
\end{equation}
and
\begin{equation}
 f_G = \left[0.1022k^2\sec(\zeta)\int_{0}^{h_\text{alt}}v^{5/3}(h)C^2_n(h)\,dh \right]^{3/5}, 
   \label{E:f_G}
\end{equation}
where $v(h)$ is the wind speed at altitude $h$~\cite{Greenwood:77,Tyson}.

For the case of moving source, $v(h)$ in Eq.~\eqref{E:f_G} is given by
\begin{equation}
\label{E:wind}
    v(h) = v_\text{wind}(h) + v_\text{app}(h) ,
\end{equation}
where $v_\text{wind}(h)$ is the natural wind speed and $v_\text{app}(h)$ is the apparent wind speed due to the movement of the source.
This assumption of simply adding two scalar speeds in Eq.~\eqref{E:wind} is justified when the moving source is mounted on a \LEO satellite because it moves at great angular speed so that $v_\text{app} \gg v_\text{wind}$. 
 We further assume that the natural wind speed follows the altitude-dependent Bufton wind profile~\cite{SasielaRichardJ2007EWPi}
\begin{equation}
\label{E:wind_natural}
    v_\text{wind}(h) = v_g + 30\exp\left[-\left(\frac{h-9400}{4800}\right)^2\right] .
\end{equation}
 Here $v_g$ is the natural wind speed and is taken to be $5$~m/s in our analysis~\cite{SasielaRichardJ2007EWPi}.

 \subsection{\LEO Satellite and Receiving End Telescope}
 \label{Subsec:LEO}
 \LEO satellite-to-ground communication is considered in this paper because of its low aperture-to-aperture loss and high speed features. 
 For simplicity, we assume that the satellite is moving in a circular orbit passing through the zenith of the detector.  Furthermore, we simply calculations by ignoring the rotation of the Earth as the orbital period of a \LEO satellite is much shorter than $1$~day.
 Then, the distance between the transmitter and receiver can be expressed as
 \begin{equation}
  \label{E:z_zeta_def}
  z(\zeta) = \sqrt{h_\text{alt}^2+2h_\text{alt}R_\oplus+R^2_\oplus\cos^2(\zeta)}-R_\oplus\cos(\zeta),
 \end{equation}
 where $R_\oplus$ is the Earth radius.
 Besides, the angular slewing rate is equal to
\begin{equation}
\label{E:slewing}
 \omega_s = \left[\frac{GM_\oplus}{h_\text{alt}^2(h_\text{alt}+R_\oplus)} \right]^{1/2}\cos^2(\zeta),
\end{equation}
 where $G$ is the universal gravitational constant and $M_\oplus$ is the Earth mass.
 Clearly, the apparent wind speed $v_\text{app}$ at height $h$ equals
 \begin{equation}
 \label{E:wind_apparent}
     v_\text{app}(h) = \omega_s h.
 \end{equation}
 To illustrate of idea, we consider the satellite moving at two different altitudes, namely, $h_\text{alt} = 400$~km and $800$~km in this paper.

 At the receiving end, the aperture coupling efficiency can be approximated by using Gaussian beam equation~\cite{Lanning21}
\begin{equation}
\label{E:eta_geo}
    \eta_\text{geo}(\omega) = 1-\exp\left[-\frac{1}{2}\frac{D^2}{\omega^2(\lambda,z(\zeta))}\right],
\end{equation}
 where $D$ is the diameter of the telescope aperture, $\lambda$ is the wavelength, $z$ is the propagation distance, 
 and the waist function $\omega$ equals
 \begin{equation}
      \omega^2(\lambda, z(\zeta)) = \omega_0^2\left[1+\frac{z(\zeta)^2}{z_R^2(\lambda)}\right], 
      \label{E:omega2}
 \end{equation}
 with $z_R = \pi\omega_0^2/\lambda$ being the Rayleigh range and $\omega_0\equiv0.7D_T/2$ being the beam waist.
 We take $D_T = 0.05$~m as the diameter of the transmitter aperture.
 The telescope parameters used are based on a real telescope in the Lulin observatory~\cite{lulin}. 
 It is a Cassegrain telescope with diameter $D = 1.03$~m, secondary mirror diameter $0.36$~m, and effective focal length $f = 8$~m.
 
 \subsection{Wavelength Selection}
 A shorter wavelength gives better quantum channel performance due to the spatial filtering strategies, geometric coupling and size of focus spot~\cite{Lanning21}.
 This conclusion is consistent with the implicit dependence of $\eta_\text{geo}$ on $\lambda$ as shown in Eqs.~\eqref{E:eta_geo} and~\eqref{E:omega2}.
 Moreover, we assume there is a field stop (\FS) in front of the signal receiver to filter the background noise.
 The size of this \FS is taken as the diffraction limit of the signal beam.
 In this configuration, the \FS can filter most of the background light while $84\%$ of the signal can pass through (if the signal is not distorted).
 Since the spot size of the beam is proportional to its wavelength, a longer wavelength increases the size of the \FS and number of background photons that pass through.
 
 Based on these factors and some site-specific conditions,  
 Gruneisen \emph{et al.} used $780$~nm as the wavelength of the signal beam and $808$~nm as the wavelength of the beacon in their daytime quantum communication experiment~\cite{Gruneisen_2021}.
 As our paper focuses more on quantum scenarios, we follow them by fixing the wavelengths of both the signal and beacon beams in our time-delayed spatial multiplexing prototype to $780$~nm.
 And for the \WDM systems that we use for performance comparison, the beacon wavelength used is set to
 $808$~nm.

\section{Reducing the Anisoplanatism by Using a Pioneer Beacon}
\label{Sec:Time}
 Anisoplanatism occurs when there is an angular difference $\theta_\text{path}$ between the beacon and the signal.
 It increases the wavefront variance $\sigma_\text{path}^2$ and hence degrades the performance of the \AO system.
 This variance can be expressed as~\cite{Tyson}
 \begin{equation}
    \sigma_\text{path}^2 = \left(\frac{\theta_\text{path}}{\theta_0}\right)^{5/3},
 \end{equation}
 where $\theta_\text{path}$ is the angle between the beacon beam path and its corresponding signal beam path, and $\theta_0$ is the isoplanatic angle calculated using Eq.~\eqref{E:theta_0}.
 There are two origins for anisoplanatism.
 Spatial anisoplanatism is induced by the spatial angular separation
\begin{equation}
 \label{E:theta_s_def}
 \theta_s = \frac{L}{z(\zeta)}
\end{equation}
 of the beams. 
 On the other hand, temporal angular separation refers to the angle between the optical paths at two different timestamps, and the time duration is the response time (in other words, the \AO lag time) $T_r$ of the system.
 The temporal angle can be expressed as
 \begin{equation}
   \label{E:theta_t_def}
   \theta_t = \omega_s T_r.
 \end{equation}
 We fix the delay time $T_r$ as the time taken between $10\%$ and $90\%$ of steady-state output.  As we model the system as an RC filter, $T_r$ can be expressed as
 \begin{equation}
  \label{E:T_r_def}
  T_r = \frac{0.35}{f_c}
 \end{equation}
 where $f_c = 500$~Hz is the $3$~dB bandwidth of the whole \AO feedback loop system.
 We further checked that this delay time $T_r$ is much shorter than $1/f_G$.
 Although our design increases the spatial angle $\theta_s$, the total anisoplanatism actually decreased as the beacon is located in front of the signal beam.
 This is because the angle that contributes to anisoplanatism is 
 \begin{equation}
     \theta_\text{path} = |\theta_t - \theta_s|.
     \label{E:angle}
 \end{equation}
 (See Fig.~\ref{Fig:time} for an illustration.)

\begin{figure}[t]
\centering
\includegraphics[width=0.48\textwidth]{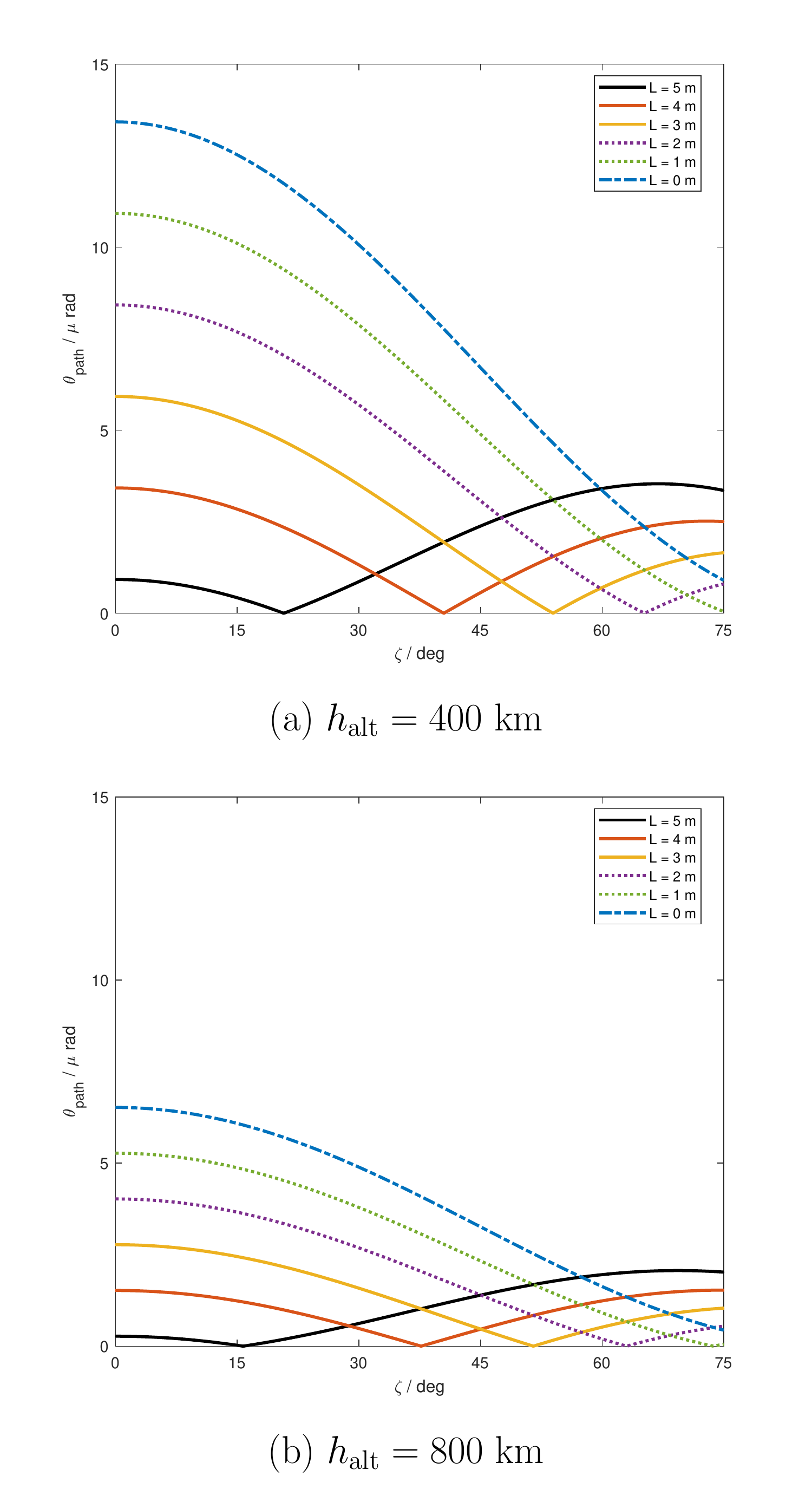}
 \caption{Total angular difference $\theta_\text{path}$ between the optical paths defined by Eq.~\eqref{E:angle}.
 The black solid, red solid, yellow solid, purple dotted, green dotted and blue dash-dotted curves are computed with spatial separation $L$ equals $5$~m, $4$~m, $3$~m, $2$~m, $1$~m and $0$~m, respectively.
 Note that the case without spatial multiplexing is the one with $L = 0$~m.
 \label{F:angle}}
\end{figure}

\begin{figure}[t]
\centering
\includegraphics[width=0.48\textwidth]{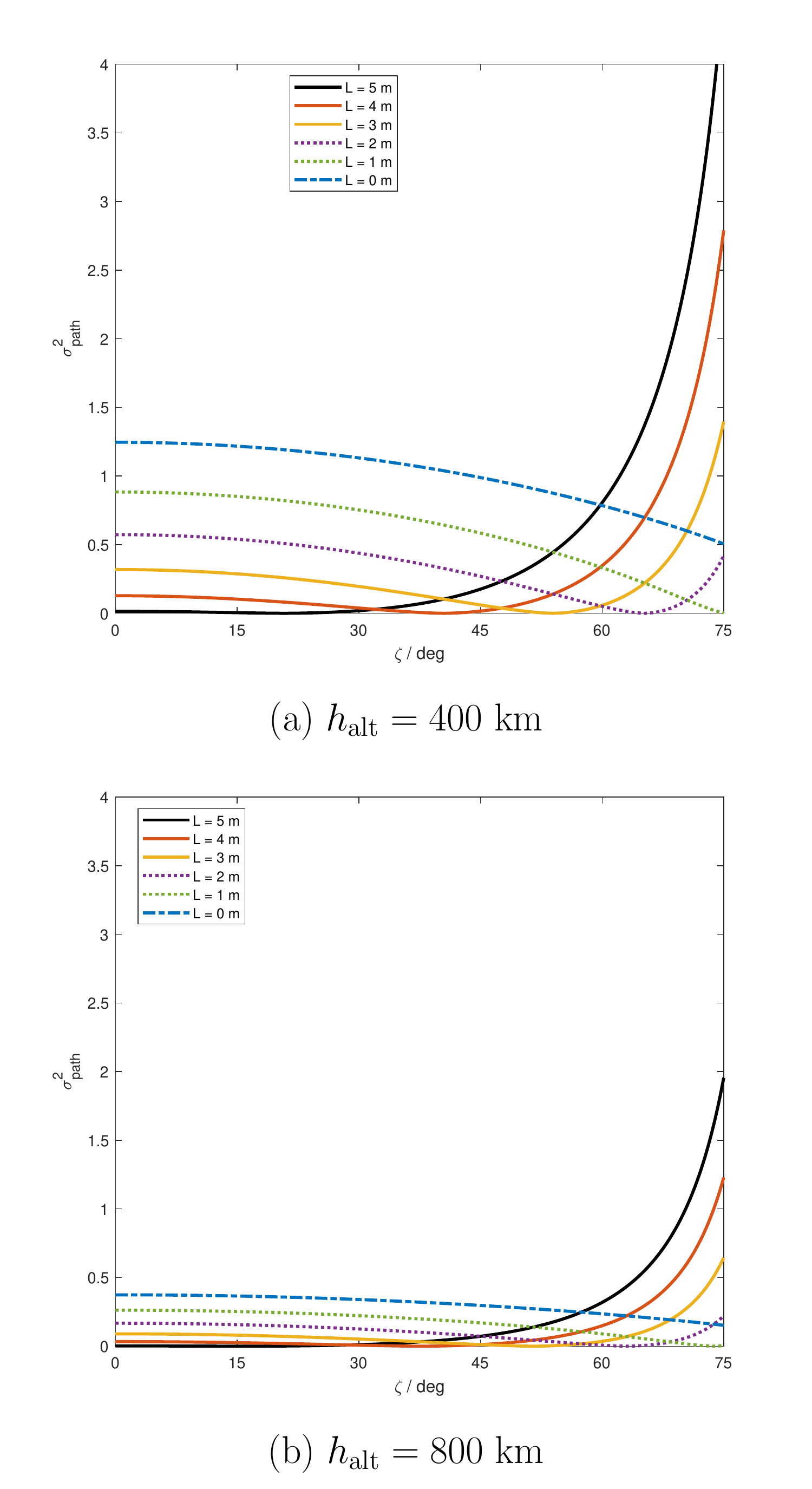}
 \caption{The wavefront variance $\sigma^2_\text{path}$ due to anisoplanatism.
 The curves are labeled in the same way as in Fig.~\ref{F:angle}.
Note that the case without spatial multiplexing is the one with $L = 0$~m.
 \label{F:sigma_path}}
\end{figure}

If the beacon and signal beams are not spatially separated, $L = 0$~m and hence $\theta_s = 0$~rad.  As shown in Figs.~\ref{F:angle} and~\ref{F:sigma_path}, spatially separating the two beams generally decreases both $\theta_\text{path}$ and $\sigma^2_\text{path}$ for zenith angle $\zeta \lesssim 60^\circ$ and $L \le 5$~m.
This demonstrates the reduction of anisoplanatism by using a pioneer beacon beam.
Moreover, from Eqs.~\eqref{E:z_zeta_def}, \eqref{E:slewing}, \eqref{E:theta_s_def} and~\eqref{E:theta_t_def}, it is easy to check that $d ( \theta_t - \theta_s) / d\zeta < 0$ for any reasonable parameters for a \LEO satellite.
No wonder why Fig.~\ref{F:angle} shows that $\theta_\text{path}$ decreases as $\zeta$ increases until reaches $0$~rad.  Beyond this point, $\theta_\text{path}$ goes up again.
That is to say, for any fixed $h_\text{alt}$, $L$ and $T_r$, there is a specific zenith angle $\zeta$ whose corresponding $\theta_\text{path}$ and hence anisoplanatism vanish.
(In reality, $\theta_\text{path} \approx 0$~rad for this set of parameters due to all the approximations made in the calculation.)
In principle, we can artificially lengthen the delay time $T_r$ or shorten the spatial separation $L$ between the beacon and signal beams to fix $\theta_\text{path}$ to its optimal value.
However, this changes the bandwidth and cross-talk calculation.
For simplicity, this kind of adjustment is not discussed in this paper.
Instead, we are going to report the effects of this type of adjustment in our followup work.
Note that even for the region that $\theta_\text{path}$ is increasing with $\zeta$, our design is still better than systems that are without spatial separation.

 To summarize, our signal pulse trick can reduce anisoplanatism as shown in Fig.~\ref{F:sigma_path}.  
 We expect this spatial separation method to be effective in obtaining a high fidelity signal through \AO.
 In Sec.~\ref{Sec:compare}, $\sigma^2_\text{path}$ is used to show that this is indeed the case.

\section{The Advantage of Spatial Multiplexing in Our Setup}
\label{Sec:Spatial}

 From the discussion of the previous section, we expect that the path angle reduction feature in our setup is compatible with \WDM and \TDM systems in the sense that
 the pioneer beam setup can be built on top of them to reduce the anisoplanatism.
 However, we can further improve the \AO system by using spatial multiplexing.
 As the beacon beam is physically separated from the signal beam, they can be distinguishable spatially.
 The setup itself is a spatial multiplexing system, which has two advantages compare to pure \WDM and \TDM systems.  First,
 the wavelength of the sources can be the same so that chromatic effects can be ignored.
 Second, there is no need to temporally interlace the signal pulses with beacon pulses.
 For our \LEO satellite setup parameters, the two beams can be spatially resolved using \AO technology because the same technology can image a satellite at $1000$~km range through a $1$~m telescope~\cite{Bennet16}.
 
 \subsection{Design Of The Beacon And Signal Beams}
 \label{Subsec:design}
 Recall that the intuition of our improved method is that two physically nearby
 light beams of similar frequency pass through more or less the same air
 column at more or less the same time should be distorted in roughly the same way.  Hence, a wavefront correction
 method based solely on the signal received by a wavefront sensing module that detects the pioneer beacon
 source beam should be able to correct both light beams at the
 same time with high fidelity.

\begin{figure}[t]
\centering
\includegraphics[width=0.25\textwidth, angle =90]{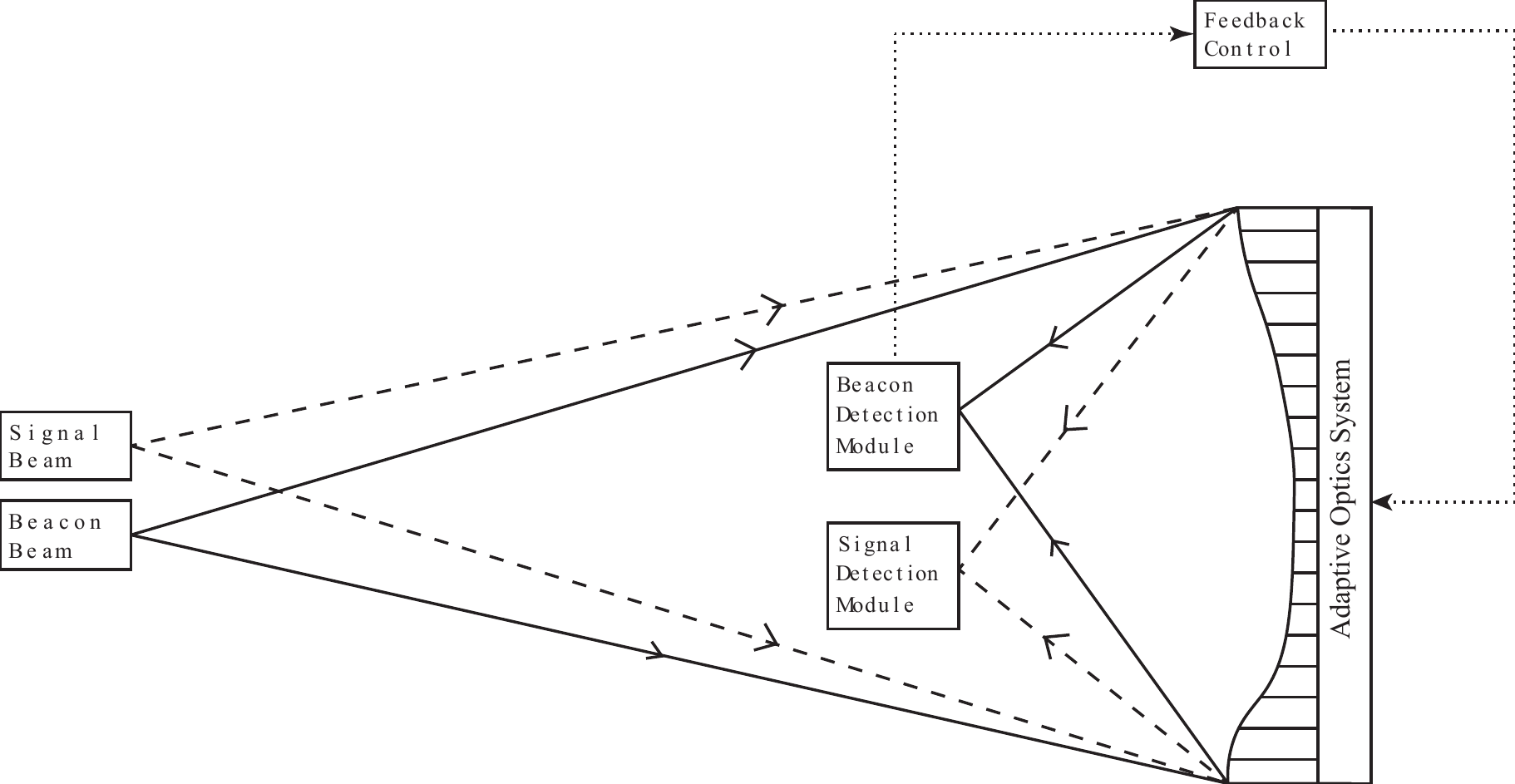}
\caption{Schematic diagram of the \AO communication system with spatially
 separated beacon and signal beams.
 \label{F:config}}
\end{figure}

 Fig.~\ref{F:config} shows the schematic of spatial multiplexing \AO system. 
 It consists of two physically nearby sources as well as
 a wavefront sensing module that detects the beacon source beam plus a nearby signal detection module
 that detects the signal source beam(s).
 To reduce photon loss in long distance communication, each of the beam source is placed at the focus of telescope on the satellite so that the emitted light beam close to the source can be well approximated by traveling plane wave.
 Our hope is that with this spatial configuration
 the optical paths of the two set of sources with the same
 or almost the same wavelength should experience more or less the same wavefront
 distortion.  The wavefront correction then goes as follows.  The beacon detection module estimates the atmospheric
 distortion and generates feedback signals to the control system.  Then
 the control system drives the actuators of the \DM in the \AO system.  This should correct the wavefront distortion of the beacon beam as
 well as the possibly much weaker signal source beam simultaneously provided that the delay time $T_r$ between the two beams is much shorter than $1/f_G$.
 Surely, in order to work, the two set of sources must be placed sufficiently far apart so that cross-talk between the beacon and signal source(s) due to effects such as
 diffraction and scattering is negligible.

 The spatial configuration of our method is similar to the standard artificial guide star
 technique used in observational astronomy~\cite{HB}.  Note, however, that
 there are two major differences.  
 First, all sources we used are artificial.  Second, our
 beacon source is placed physically closed (and not just close in terms of
 apparent angular separation) to the signal source(s).
 We remark that this spatial configuration works not just for secure quantum communications.  It is
 directly applicable to classical optical communication in free-space as well.
 And in this case, the intensity of the signal source(s) need not be low.
 In addition, our method is applicable to ground-based, air-to-ground as well
 as satellite-to-ground communications, stationary as well as moving sources
 relative to the sensing and detecting modules.
 Furthermore, a nice feature of our method is that the signal transmission rate will then be independent of
 the beacon source.

\subsection{Minimum Physical Distance Between The Beacon And Signal Sources}
\label{Subsec:Min_Source_Distance}
 The minimum possible distance between the beacon and signal sources is determined by
 both the resolving power of the optics and the level of cross-talk between the
 two set of sources.  Note that upon successful \AO correction, the center of
 the image of the beacon beam should be around the center of the optically
 sensitive surface of the wavefront sensing module.  We put a field stop
 in the signal detection module to filter the noise spatially.
 Naturally, we set the radius of the field stop to the 
 diffraction limit of the signal detection module~\cite{Gruneisen2014}.
 The diffraction pattern depends on the structure of the telescope.
 In our concrete illustration, we use a 1.03~m Cassegrain telescope whose parameters are taken from a real telescope in Lulin Observatory~\cite{lulin}.
 The light intensity of the beacon beam at a distance $x$ away from the center equals
 \begin{equation}
  I_\text{R}(x) \approx I_\text{R}(0) \left( \frac{f\lambda}{\pi D x}
  \right)^2 \left[ J_1 \left( \frac{\pi D x}{f\lambda} \right) - b J_1 \left(
  \frac{b \pi D x}{f\lambda} \right) \right]^2 ,
  \label{E:intensity}
 \end{equation}
 where $f$ is the effective local length of the telescope, $b = 0.36/1.03$ is
 ratio of the diameters of the secondary to primary mirrors of the Cassegrain
 telescope used, $I_\text{R}(0)
 \approx 2 \epsilon_0 c {\mathcal E}_\text{R}^2 \pi^2 (D/2)^4 / R^2$, and
 $J_1(\cdot)$ is the order one Bessel function of the first kind.  Hence,
 the total light energy flux of the beacon beam imparted on the optically
 sensitive surface of the signal detection module is $\iint_\text{\FS}
 I_\text{R}(x) \,dA$ where the integral is over the area of the field stop
 of the signal detection module. For example, when $L = 2$~m, $\iint_\text{\FS}
 I_\text{R}(x) \,dA = 4.36 \times 10^{-15} I_\text{R}(0)$. The minimum distance should be set according to the
 required decay from the beam center. Otherwise, stray beacon beam photons will seriously
 affect the signal detection statistics.  

\subsection{Scattering Noise by the Strong Beacon Beam}
\label{Subsec:Scattering}
 The scattering caused by the strong beacon beam will affect the background noise of the system and hence in satellite-to-ground \QKD application the final secret key rate.
 Some photons from the beacon may enter the signal receiving module and create errors.
 Here we estimate the scattering by the strong laser in the clear sky scenario.
 We use sky-scattering noise to get a rough estimate on the laser scattering
 noise. The equation for calculating the number of sky-noise photons entering the system
 is given by~\cite{Gruneisen2014},
\begin{equation}\label{E:scattering}
 N_b = \frac{H_b(\lambda)\Omega_\text{FOV} \pi D_{R}^{2}\lambda\Delta\lambda\Delta t}{4hc},
\end{equation}
 with $H_b$ in W\,m$^{-2}$\,sr\,$\mu$m is the sky radiance,
 $\Omega_\text{FOV} = \pi \Delta \theta^2 /4$ is the solid-angle field
 of view with a field stop, $D_R$ is diameter of the receiver
 primary optic, $\Delta\lambda$ equals to the spectral filter bandpass
 in $\mu$m, and $\Delta t$ is the photon integration time of the
 receiver measured in meters.
 Furthermore, $\Delta \theta$ is calculated by $D_\text{\FS}/f$ with $D_\text{\FS}$ being the
 diameter of the field stop. We assume $\Delta \lambda = 1$~$\mu$m as both
 beams use the same or nearly the same wavelength, the 
 spectral filter is not able to block the photons from the beacon beam.

 In astrophotography, a bright star that is close to target can be used as
 a beacon to probe the channel. Therefore, the brightness of the beacon laser
 should be similar to a bright star. The sky radiance caused by the laser can be 
 estimated by the sky radiance by the stars.
 Typical sky radiance is about $1.5\times 10^{-5}$~W\,m$^{-2}$\,sr\,$\mu$m under moonless clear night condition~\cite{Er_long_2005}. Using the
 parameters mentioned above and let $\Delta t = 1$~ns,
 the probability of receiving a beacon photon will be in the order of $10^{-8}$, which is good enough in practice.

\section{Comparison Of Performance Of Our Prototype With Those Using Pure \TDM and \WDM}
\label{Sec:compare}
\subsection{The Strehl Ratio}
\label{Subsec:Strehl}
The Strehl ratio $S$ is a well-known metric to determine the turbulence strength and performance of optical systems.
It is defined as the ratio of the peak intensity of a distorted beam spot and the peak intensity of the beam with no distortion.
If the Strehl ratio equals to one, the wavefront is not aberrated.
Without using \AO, the Strehl ratio of the signal is~\cite{Tyson}
\begin{equation}
    S_\text{aber} = \left[1+\left(\frac{D}{r_0}\right)^{5/3}\right]^{-6/5}.
\end{equation}
When \AO is used, the performance of the system can be estimated by~\cite{Tyson}
\begin{equation}
    S_\text{i} = \exp(-\sigma_\text{i}^2),
\end{equation}
where $S_\text{i}$ is the Strehl ratio of system $i$, and $\sigma_i^2$ is the corresponding wavefront variance, which leads to system performance degradation.
Here we compare three systems, namely, those using time-delayed spatial multiplexing, pure \TDM, and pure \WDM.
Their wavefront variance can be written as~\cite{Tyson,Devaney:08}
\begin{equation}
    \sigma^2_\text{SS} = \sigma_\text{band}^2+\sigma_\text{path, SS}^2,
\end{equation}
\begin{equation}
    \sigma^2_\text{TDM} = \sigma_\text{band}^2+\sigma_\text{path, $\overline{\text{SS}}$}^2,
\end{equation}
and
\begin{equation}
\label{E:sigma_WDM}
    \sigma^2_\text{WDM} = \sigma_\text{band}^2+\sigma_\text{path, $\overline{\text{SS}}$}^2+\sigma_\text{d}^2+\sigma_\text{ch}^2+\sigma_\phi^2,
\end{equation}
where SS ($\overline{\text{SS}}$) indicates that the system is (is not) using the pioneer beacon setup, 
and the descriptions of the $\sigma^2$'s are as follows
\begin{itemize}
  \item \makebox[0.85cm][l]{$\sigma_\text{band}^2$}: bandwidth limitation induced wavefront variance;
  \item \makebox[0.85cm][l]{$\sigma_\text{path}^2$}: temporal and spatial anisoplanatism induced wavefront variance;
  \item \makebox[0.85cm][l]{$\sigma_\text{d}^2$}: chromatic effect on the diffraction pattern induced wavefront variance;
  \item \makebox[0.85cm][l]{$\sigma_\text{ch}^2$}: chromatic path length error induced wavefront variance; and
  \item \makebox[0.85cm][l]{$\sigma_\phi^2$}: chromatic anisoplanatism induced wavefront.
\end{itemize}
Details of the calculations and expressions of these $\sigma^2$'s can be found in Appendix~\ref{Sec:sigma} from Eq.~\eqref{E:App1_begin} to Eq.~\eqref{E:App1_end}.
Note that $v_\text{app}(h)$ cannot be reduced by using the pioneer beacon setup.
The \AO system still ``sees'' a fast-changing beacon. 
This is reflected in the wavefront variance due to bandwidth $\sigma_\text{band}^2$.
The significance of our design is the improvement on $\sigma_\text{path}^2$
As the aim of this Subsection is to compare different multiplexing methods,
we ignore system degradation due to factors that are not related to temporal, chromatic or anisoplanatic effects in our calculation.
Furthermore, for simplicity, we do not take interlacing into account for all \TDM calculations in this paper.

\begin{figure}[t]
\centering
\includegraphics[width=0.48\textwidth]{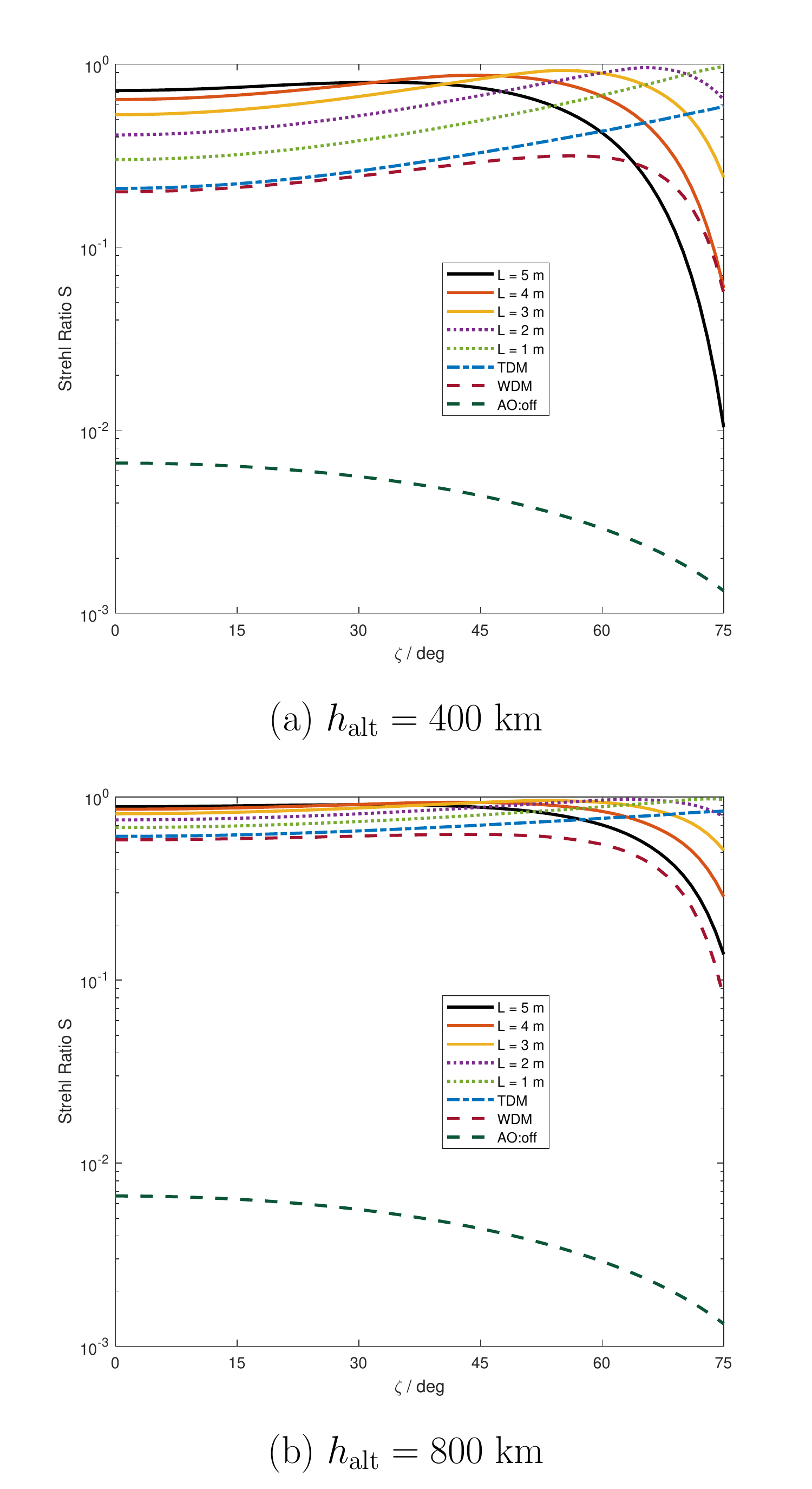}
\caption{Strehl ratio $S$ of different system configurations as a function of zenith angle $\zeta$.
 The black solid, red solid, yellow solid, purple dotted, green dotted curves are computed with $D = 1.03$~m and spatial separation $L$ equals $5$~m, $4$~m, $3$~m, $2$~m and $1$~m, respectively.  The blue dash-dotted and brown dashed curves are for pure \TDM and \WDM methods, respectively.  The dark green dashed curve is for turning off \AO.
 \label{F:Strehl}}
\end{figure}

Fig.~\ref{F:Strehl} shows the results of the calculation.
For both $h_\text{alt} = 400$~km and $800$~km, the Strehl ratio of the \TDM system is higher than that of the \WDM system which in turn is much higher than when \AO is turned off for all zenith angles $\zeta$.
More importantly, for $h_\text{alt} = 400~$km, separating the beacon and the signal beams up to $L = 5$~m gives higher $S$ than the \TDM system when $\zeta \lesssim 60^\circ$.  And for $h_\text{alt} = 800$~km, separating the two beams up to $L = 5$~m always gives a higher $S$ than the \TDM system when $\zeta \lesssim 55^\circ$.
This means that anisoplanatism is the dominant factor when calculating the Strehl ratio.

\begin{figure}[t]
\centering
\includegraphics[width=0.48\textwidth]{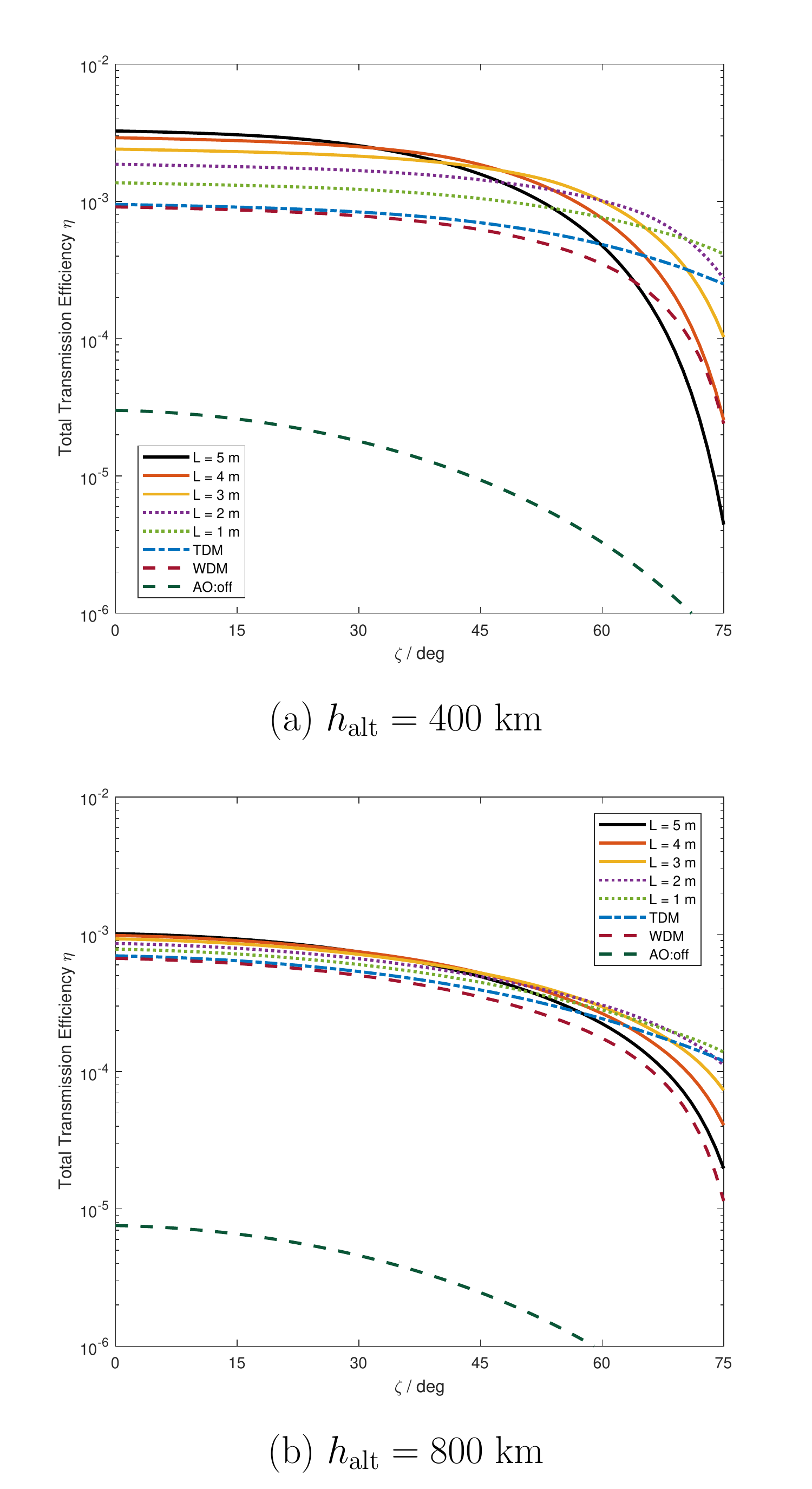}
 \caption{Total transmission efficiency $\eta$ of different system configurations as a function of zenith angle $\zeta$.
 The curves are labeled in the same way as in Fig.~\ref{F:Strehl}.
 \label{F:eta}}
\end{figure}

\subsection{Channel Efficiency}
\label{Subsec:eta}
The total efficiency $\eta$ of the satellite-to-ground channel, which depends on the atmospheric condition, can be expressed as~\cite{Lanning21, GruneisenMarkT2017Msqc}
\begin{equation}
    \eta = \eta_\text{trans}(\zeta)\eta_\text{rec}\eta_\text{spec}\eta_\text{det}\eta_\text{geo}(\zeta)\eta_\text{\FS}(\zeta) .
    \label{E:eta_def}
\end{equation}
Here factors that implicit or explicit depend on the zenith angle $\zeta$ are emphasized by explicitly showing this dependence.
In Eq.~\eqref{E:eta_def}, $\eta_\text{trans}(\zeta)$ is the free-space transmission efficiency.  It depends on the zenith angle $\zeta$ although the dependence is rather weak.
To simplify matter, we assume that $\eta_\text{trans}$ is a linear function with of $\zeta$ with $\eta_\text{trans} = 0.92$ at $\zeta = 0^\circ$ and $\eta_\text{trans} = 0.74$ at $\zeta = 75^\circ$.  These two values are the results obtained by Gruneisen \emph{et al.} in their the MODTRAN simulation for clear sky conditions~\cite{GruneisenMarkT2017Msqc}.
Actually, we have tried a few variations on $\eta_\text{trans}$ and found that it does not change our results in any significant way.
As for the other factors in Eq.~\eqref{E:eta_def}, we follow Lanning \emph{et al.}~\cite{Lanning21} by picking the efficiency of the receiver $\eta_\text{rec} = 0.5$,
the efficiency of the spectral filter $\eta_\text{spec} = 0.9$,
the detector efficiency $\eta_\text{det} = 0.8$.
Besides, the aperture-to-aperture coupling efficiency $\eta_\text{geo}(\zeta)$ is
given by Eq.~\eqref{E:eta_geo}, and
the efficiency of the \FS is given by~\cite{Lanning21}
\begin{equation}
\label{eq:eta_FS}
    \eta_\text{\FS}(\zeta) = 0.84S.
\end{equation}
Note that $\eta_\text{\FS}$ implicitly depends on the zenith angle $\zeta$ through the Strehl ratio $S$.

We find from Fig.~\ref{F:eta} that the total transmission efficiency $\eta$ always decreases as zenith angle $\zeta$ increases.  This is what we expect.
Note further that $\eta_\text{\FS}$ is the only factor that is related to the distortion loss in this framework.
From Eq.~\eqref{eq:eta_FS},
a higher Strehl ratio $S$ implies a higher $\eta$.
Thus, Fig.~\ref{F:eta} shows that the aperture coupling efficiency plays a significant role as the altitude increases.
Higher altitude can decrease the anisoplanatism, but the total channel transmission efficiency $\eta$ decreases due to larger the beam size spread.
Fig.~\ref{F:eta} also depicts that for $L \le 5$~m, $\eta$ increases with $L$ when the zenith angle $\zeta \lesssim 30^\circ$.

\section{Application in Quantum Key Distribution}
\label{Sec:QKD}
We now consider the effect of cross-talk between the beacon and signal beams.
To analyze the effectiveness of our protocol, we choose the most extreme
setting that the signal beam is a weak coherent photon source used in decoy state \BB \QKD using three photon intensities~\cite{decoy1,decoy2,Ma_2005}, namely, the vacuum source and two phase randomized Poissonian distributed sources with intensities $\mu$ and $\nu$.
In this setting, cross-talk noise could affect the system seriously by increasing the quantum bit error rate (\QBER) and hence lowering secret key rate.
In this regard, if our protocol works better than existing satellite-to-ground
\QKD setups, then it should also work in practically all realistic satellite-to-ground
communication, both classical and quantum.

\begin{table}[t]
\centering
\begin{tabular}{lcl}
\hline
\multicolumn{3}{c}{QKD parameters}                                                      \\ \hline
Quantity                 & Symbol            & Value                              \\ \hline
Signal-state mean photon numbers      & $\mu$             & $0.7$                              \\
Decoy-state mean photon numbers             & $\nu$             & $0.1$                        \\
Repetition rate          & $f_\text{source}$ & $10$~MHz                  \\
Sky radiance             & $H_b$             & $25$~Wm$^{-2}$sr\,$\mu$m \\
Dark count rate          & $f_\text{dark}$   & $10$~Hz                   \\
Polarization cross-talk  & $e_d$             & $0.01$ \\
System noise error  & $e_0$             & $0.5$  \\
Spectral filter bandpass & $\Delta \lambda$  & $0.2$~nm                   \\
Detection time           & $\Delta t$        & $1$~ns \\                        
Error-correction efficiency          & $f_e$        & $1.22$  \\
\hline
\end{tabular}
 \caption{Parameters used for calculating the secret key rate of decoy state \BB protocol used by Lanning \emph{et al.}~\cite{Lanning21}.
 \label{table:key_rate_parameters}}
\end{table}

Recall that the background detection probability can be written as
\begin{equation}
    Y_0 = (N_b+N_\text{cross})\eta_\text{spec}\eta_\text{rec}\eta_\text{det}+4f_\text{dark}\Delta t,
\end{equation}
where $N_b$, $N_\text{cross}$, $f_\text{dark}$ and $\Delta t$ are the sky photon noise, cross-talk noise due to the beacon, the dark count rate of the detectors and the detection time window, respectively.
Moreover, $N_b$ is calculated using Eq.~\eqref{E:scattering} with the parameters stated on Table~\ref{table:key_rate_parameters}.
As a conservative estimate, we assume that $N_\text{cross}$ is 100~times of the scattering noise calculated in Sec.~\ref{Subsec:Scattering} when spatial multiplexing is used.
Furthermore, for \WDM and \TDM systems, we take the liberty to set $N_\text{cross} = 0$.
The rest of the calculations are well known and can be found in Ma \emph{et al.}~\cite{Ma_2005}.  We include them here for readers' convenience.
The \QBER can be expressed as
\begin{equation}
    E_\mu = \frac{e_0Y_0+e_d(1-e^{-\eta\mu})}{Y_0+1-e^{-\eta\mu}},
\end{equation}
where $e_0$ and $e_d$ are the system noise error rate and  polarization cross-talk.
The value of the parameters are based on those used by Lanning \emph{et al.}~\cite{Lanning21} and are presented in Table~\ref{table:key_rate_parameters}.
These parameters are optimized to give the highest possible secret key rate for free-space photon
transmission in the so-called asymptotic limit, namely, for the case of an arbitrarily large number of photon transfer.
The secret key rate (more accurately, a provably secure lower bound of the number of secret key obtained at the end divided by the number of signal photon pulses emitted by the satellite) can be written as
\begin{equation}
 R\geq q\{e^{-\mu}\mu Y_1[1-h_2(e_1)]-f_e Q_\mu h_2(E_\mu)\},
 \label{E:rate_equation}
\end{equation}
where $h_2(x)$ is the binary entropy function, $Y_1$ is the single photon state yield 
\begin{equation}
        Y_1 = \frac{\mu}{\mu\nu-\nu^2}\left(Q_\nu e^\nu - Q_\mu e^\mu\frac{\nu^2}{\mu^2}-\frac{\mu^2-\nu^2}{\mu^2}Y_0\right),
\end{equation}
$e_1$ is the single photon state error rate
\begin{equation}
    e_1 =\frac{Q_\nu E_\nu e^\nu -e_0 Y_0}{Y_1 \nu},
\end{equation}
and $Q_\mu$ is the gain at intensity $\mu$
\begin{equation}
    Q_\mu = Y_0 + 1 - e^{-\eta \mu}.
\end{equation}

Last but not least, $q$ is the probability that the trusted agents on the satellite and on the ground use the same basis for preparing and measuring their signal photons in their \QKD experiment.
In the work of Lanning \emph{et al.}~\cite{Lanning21}, $q$ is chosen to be $1/2$.  But in the asymptotic
limit, the optimized key rate can be computed by taking the limit of $q\to 1$
using biased bases selection~\cite{biased_bases}.
Note however that as $q$ is a parameter independent of the channel and the \AO setup used.  It only appears as a multiplication factor in the R.H.S. of Eq.~\eqref{E:rate_equation} as far as the key rate is concerned.
Consequently, if the key rate of a certain method is higher than that of another method for a fixed $q > 0$, then the key rate of the former method is always higher than the later for all $q \in (0,1]$.
Therefore, we only need to compare the provably secure key rates of different methods by fixing, say, $q = 1/2$.

\begin{figure}[t]
\centering
\includegraphics[width=0.48\textwidth]{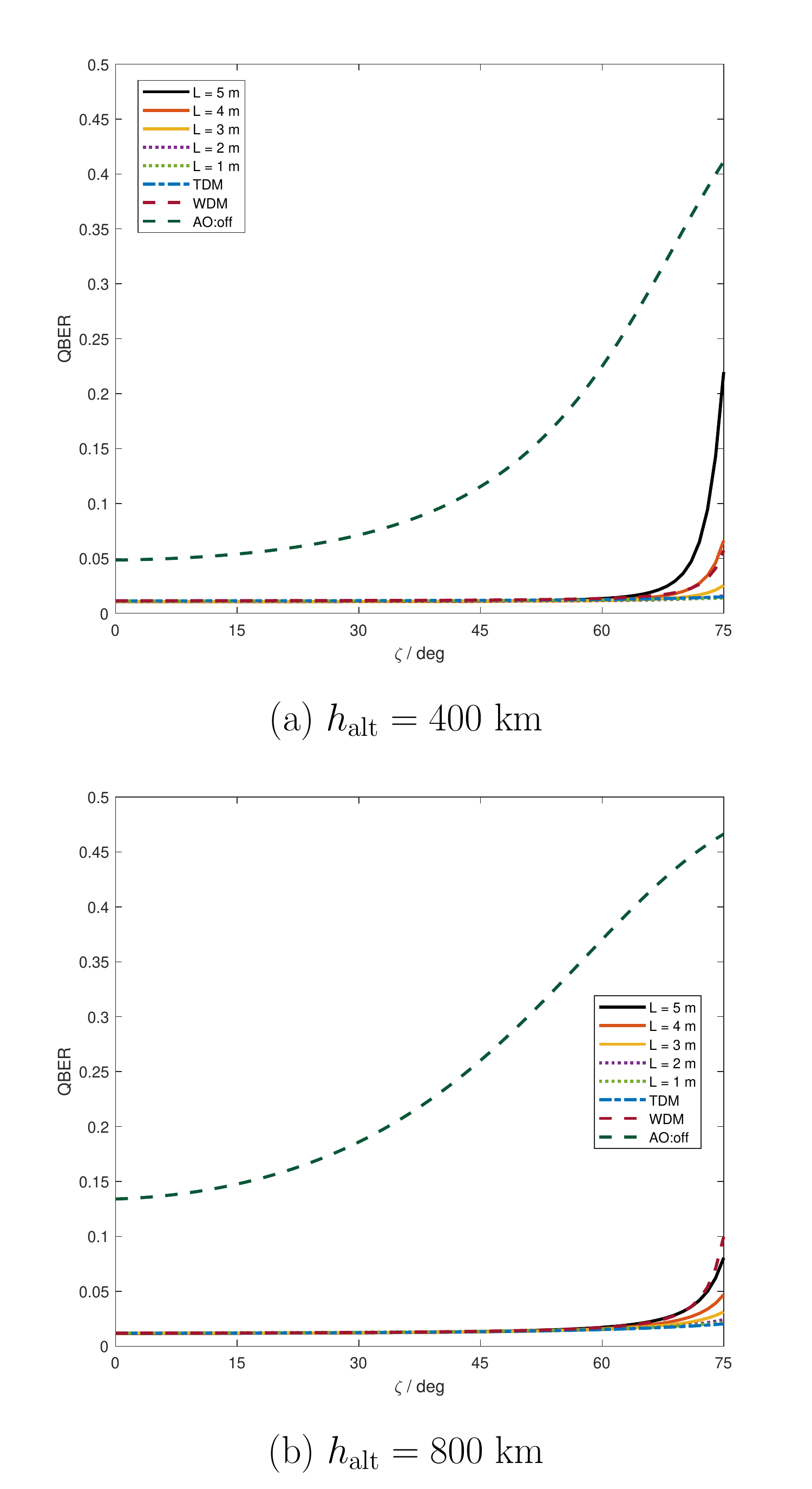}
 \caption{\QBER{s} of different system configurations.
 The curves are labeled in the same way as in Fig.~\ref{F:Strehl}.
 \label{F:QBER}}
\end{figure}

\begin{figure}[th]
\centering
\includegraphics[width=0.48\textwidth]{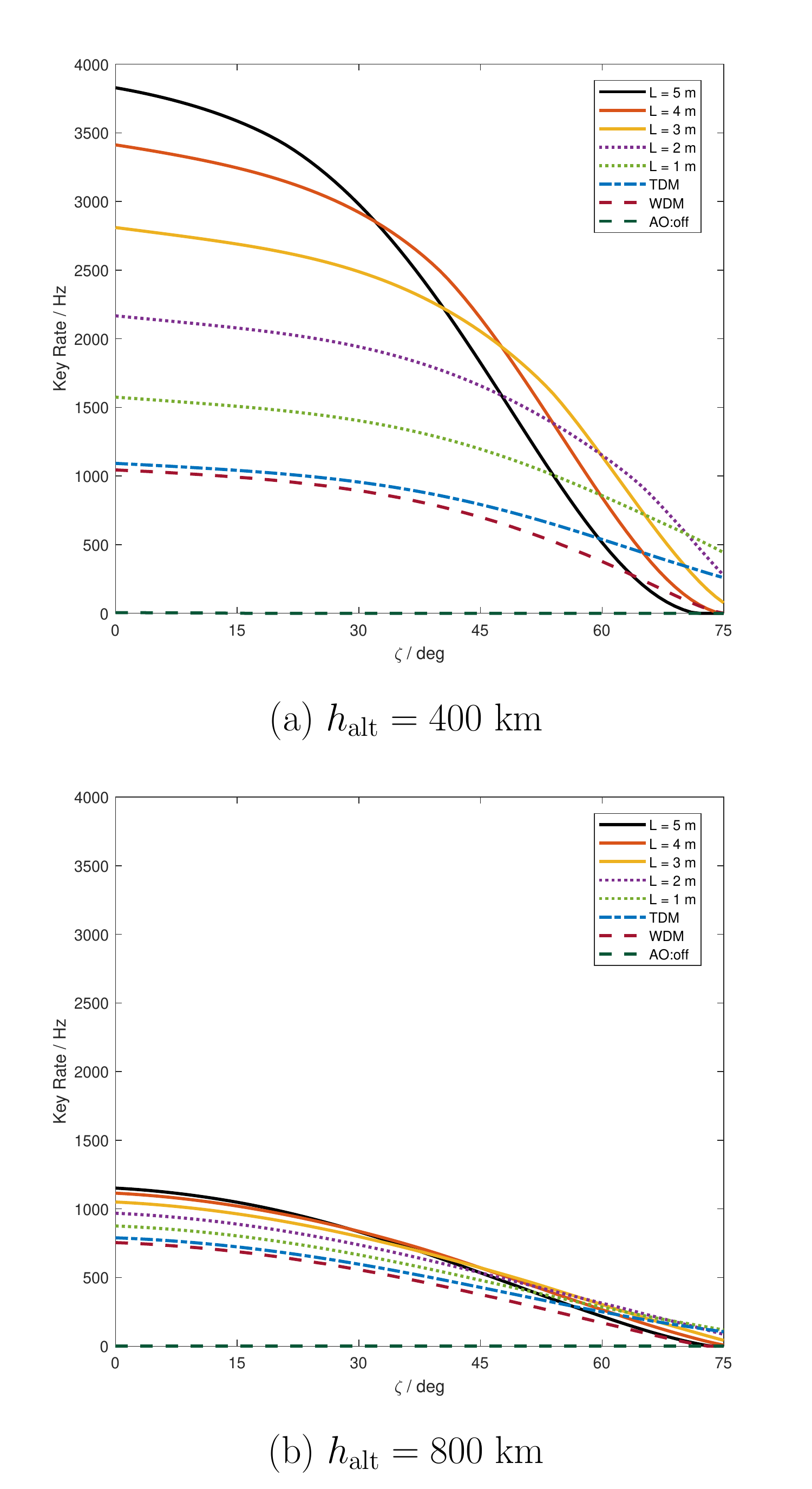}
\caption{Provably secure key rates of different system configurations.
 The curves are labeled in the same way as in Fig.~\ref{F:Strehl}.
 Note that the key rate of pure \WDM and pure \TDM systems is $0$~Hz when $h_\text{alt} = 400$~km while the key rate is close to $0$~Hz when \AO is not used and $\zeta$ is small.
 \label{F:key-rate}}
\end{figure}

While theorists use a dimensionless key rate like the one in Eq.~\eqref{E:rate_equation} as one of the effectiveness metric to study \QKD protocols, from a practical point of view, here we use the ``experimentalist'' version of the key rate, namely, $f_\text{source} R$ in this study.
It tells us the lower bound of the number of provably secure secret key bits generated per unit time.
Figs.~\ref{F:QBER} and~\ref{F:key-rate} show the final results of the \QBER and the ``experimentalist'' version of the key rate.
By comparing Fig.~\ref{F:Strehl} with Fig.~\ref{F:QBER}, we find that \QBER is anti-correlated with $S$.  This is not surprising as high aberration is likely to cause higher detection error.
In fact, our result suggests that aberration is the main source of quantum bit error in our system.  In other words, the high \QBER is caused by low signal to noise ratio as more and more signal detected is from noise.
For the key rate, Fig.~\ref{F:key-rate} depicts that the curves of the \AO systems have the same pattern as the total channel efficiency for $\zeta \lesssim 65^\circ$.  This is because the \QBER is more or less a constant for a given $L$.
Furthermore, Fig.~\ref{F:key-rate} shows that \QKD is not possible by turning off \AO.
More importantly, for $\zeta \lesssim 55^\circ$, the key rate of using pioneer beacon beam plus \AO is higher than that of the \TDM system which in turn is higher than that of the \WDM system.  Comparing to the \TDM system, at altitude $h_\text{alt} = 400$~km and zenith angle $\zeta \le 30^\circ$, the improvements are at least $215\%$ and $47\%$ for $L = 5$~m and $1$~m, respectively.  Whereas for $h_\text{alt} = 800$~km, the corresponding improvements are $40\%$ and $11\%$, respectively.
These improvement figures are computed by setting the response time $T_r$, which is inversely proportional to the bandwidth of the system.  For systems with lower bandwidth, there are more rooms for improvement as the temporal angle is larger.  We can further increase $L$ to obtain a more significant key rate improvement.

\begin{figure}[t]
\centering
\includegraphics[width=0.48\textwidth]{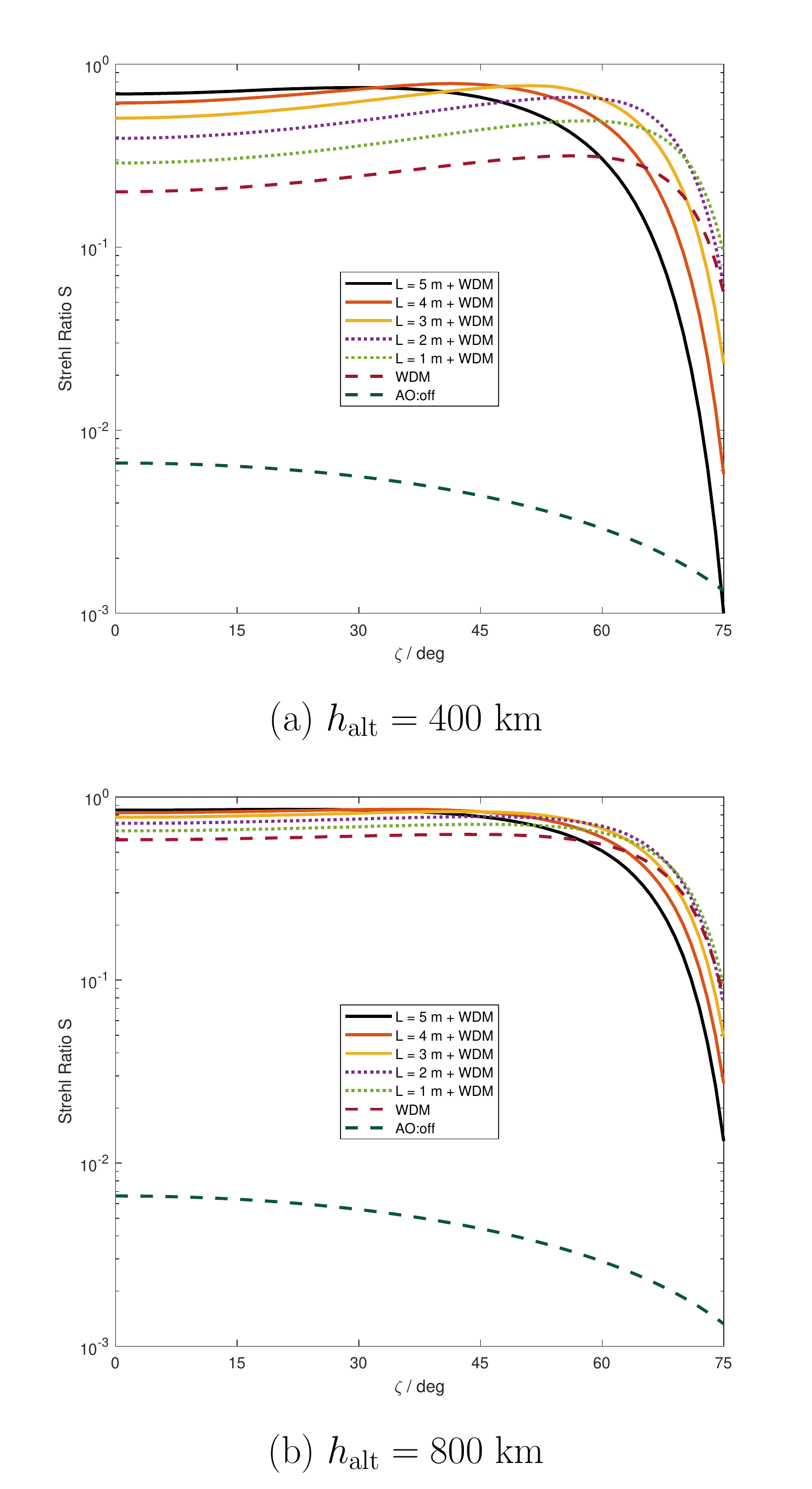}
\caption{Strehl ratio of \WDM systems. 
 Similar to the labeling scheme in Fig.~\ref{F:Strehl},
 the black solid, red solid, yellow solid, purple dotted, green dotted curves are computed using $D = 1.03$~m and \WDM with spatial separation $L$ equals $5$~m, $4$~m, $3$~m, $2$~m and $1$~m, respectively.  The brown dashed curve is for pure \WDM method.  Lastly, the dark green dashed curve is for turning off \AO.
 \label{F:Strehl_WDM}}
\end{figure}

\begin{figure}[ht]
\centering
\includegraphics[width=0.48\textwidth]{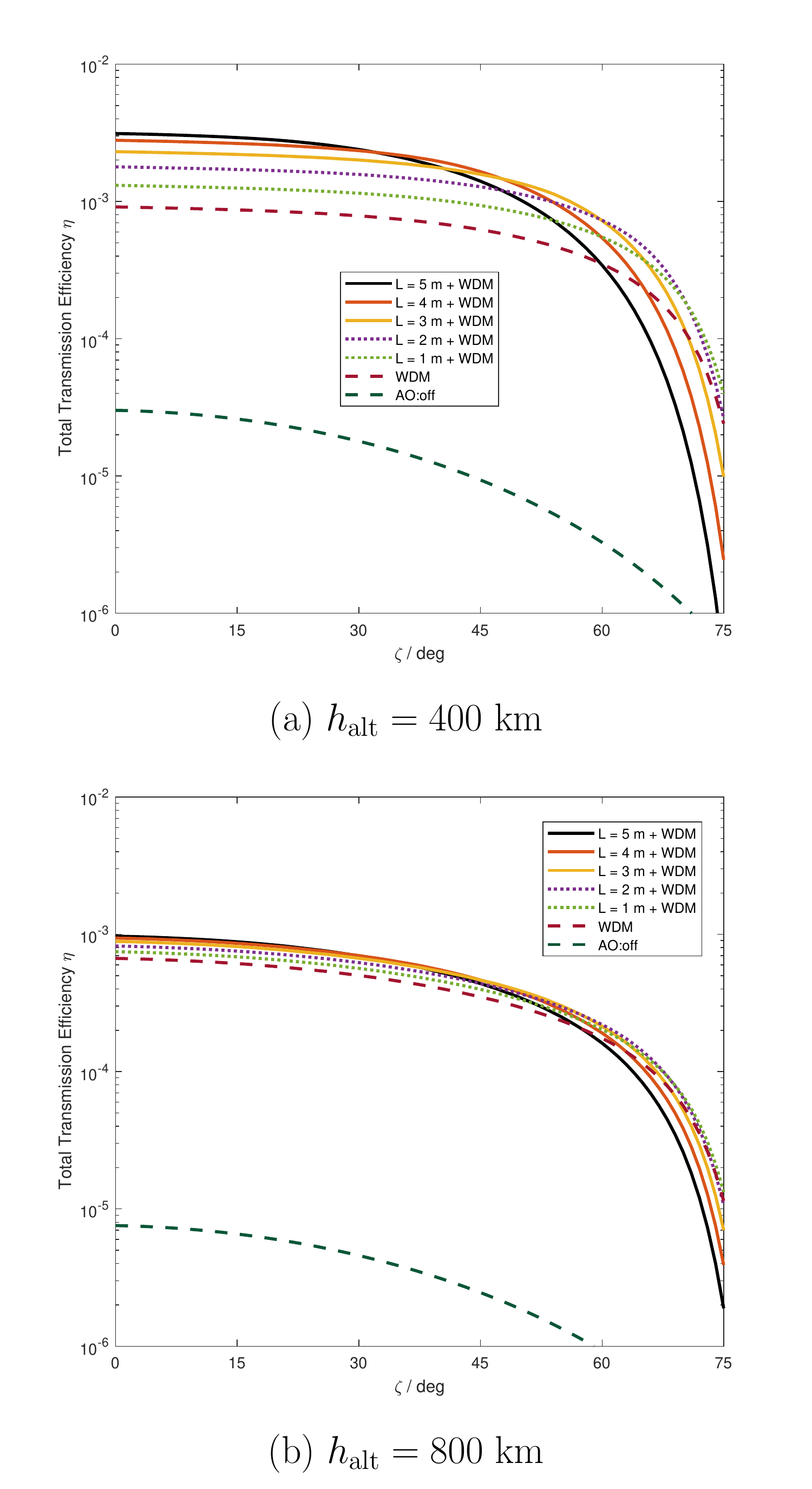}
\caption{Total transmission efficiency $\eta$ of \WDM systems.
 The curves are labeled in the same way as in Fig.~\ref{F:Strehl_WDM}.
 \label{F:eta_WDM}}
\end{figure}

\begin{figure}[ht]
\centering
\includegraphics[width=0.48\textwidth]{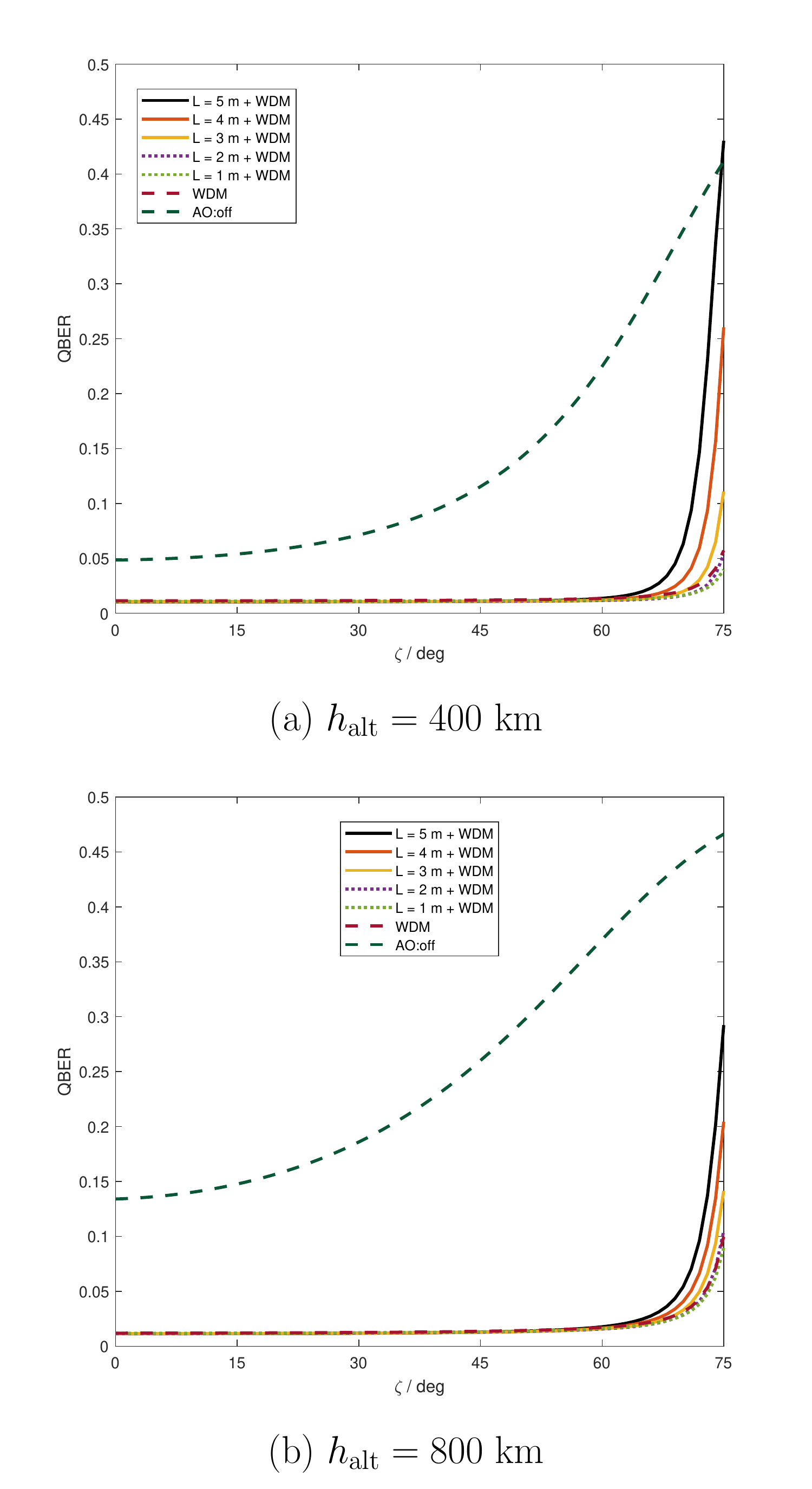}
\caption{\QBER of \WDM systems. 
 The curves are labeled in the same way as in Fig.~\ref{F:Strehl_WDM}.
 \label{F:QBER_WDM}}
\end{figure}

\begin{figure}[ht]
\centering
\includegraphics[width=0.48\textwidth]{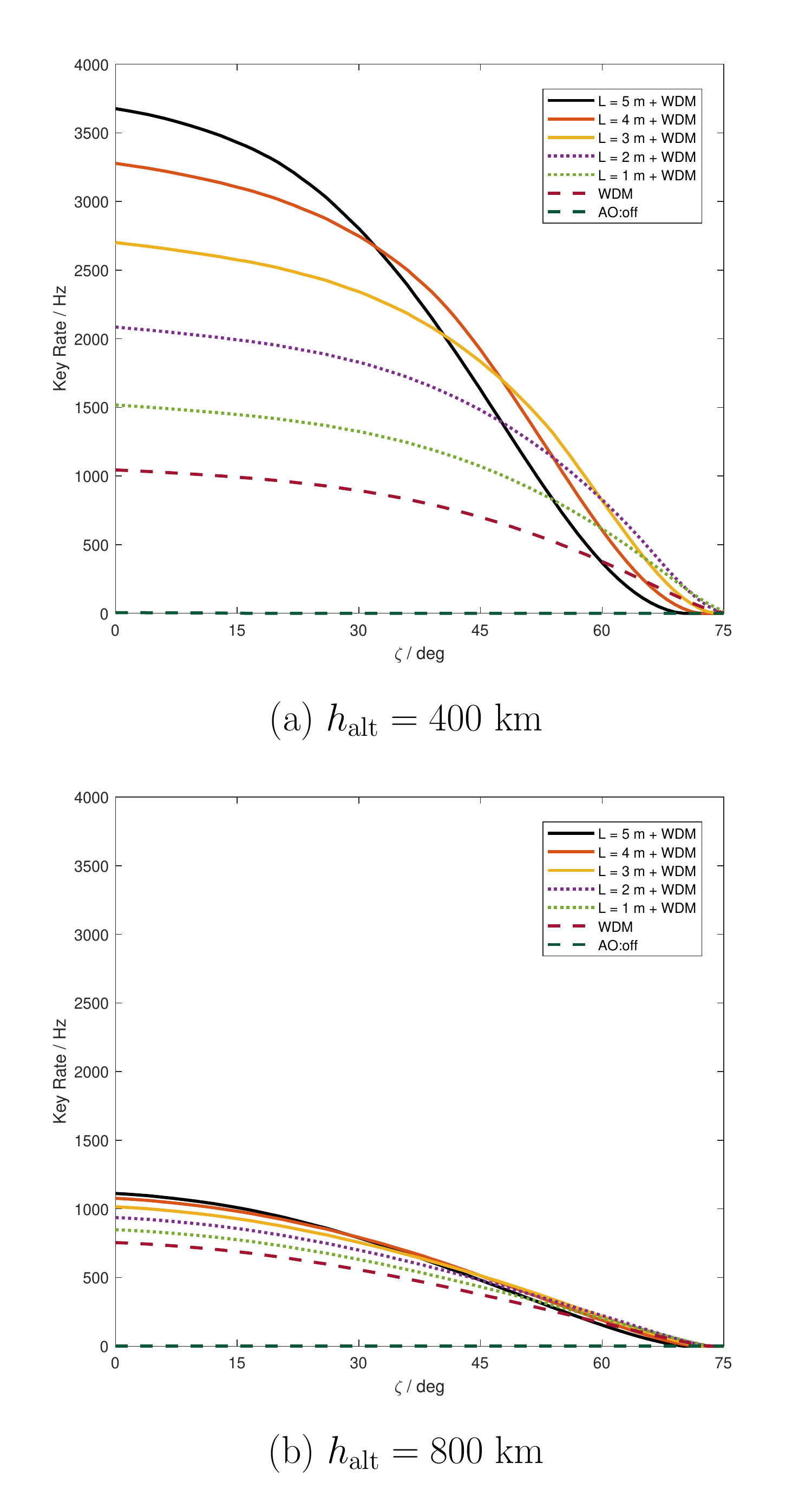}
\caption{Key rate of \WDM systems.
 The curves are labeled in the same way as in Fig.~\ref{F:Strehl_WDM}.
 Note that when \AO is off, the key rate is $0$~Hz except when $h_\text{alt} = 400$~km and $\zeta$ is low.
 \label{F:R_WDM}}
\end{figure}

\section{Improving \WDM systems using a Pioneer Beacon}
\label{Sec:WDM_improve}
As \WDM is a popular method to combine the beacon and signal beams, it is easier to upgrade the system with a pioneer beacon than building a spatial multiplexing system.
Here we present the improvement to \WDM systems with our idea.
To study that, we modify Eq.~\eqref{E:sigma_WDM} to~\cite{Tyson}
\begin{equation}
 \sigma^2_\text{WDM} = \sigma_\text{band}^2+\sigma_\text{path,\text{SS}}^2+\sigma_\text{d}^2+\sigma_\text{ch}^2+\sigma_\phi^2.
\end{equation}

Using the same set of parameters and following the same analysis in Secs.~\ref{Sec:compare} and~\ref{Sec:QKD}, the Strehl ratio, \QBER, and key rate are shown in Figs.~\ref{F:Strehl_WDM}--\ref{F:R_WDM}, respectively.
We remark that the curves, including their shapes and trends, in these figures are very similar to those in Figs.~\ref{F:Strehl}--\ref{F:key-rate}.  In other words, the performance of our improvement to the \WDM system is similar to that of the spatial multiplexing one.

In summary, upgrading a pure \WDM system with a pioneer beacon beam can significantly increase the secret key rate in decoy-state \QKD using phase-randomized Poissonian source.
Furthermore, by comparing Fig.~\ref{F:key-rate} with Fig.~\ref{F:R_WDM}, for the same value of $L$, the two systems have comparable key rates for $\zeta < 45^\circ$ --- the latter is at least 90\% of the former.
Again, these improvement figures are computed by setting the response time $T_r$, which is inversely proportional to the bandwidth of the system.  For systems with lower bandwidth, there are more rooms for improvement as the temporal angle is larger.  We can further increase $L$ to obtain a more significant key rate improvement.
In summary, our results imply that upgrading existing \WDM systems is a very attractive alternative to building entirely new spatial multiplexing ones.

\section{Conclusions}
\label{Sec:conclusion}
In this paper, we report a method to apply \AO technologies
to optical communication systems. The main ideas are the
spatial separation of the beacon and the signal beam.
For fast-moving sources, our design, which in essence performs time-based \AO pre-compensation, can reduce the angle between the optical paths of the beacon and the corresponding signal.
Thus, it can reduce the anisoplanatism of the \AO correction. 
We estimate the cross-talk caused by the diffraction and 
the scattering of the beacon. As there is a field stop in the beacon 
receiving module and the power of the beacon is not high, the 
cross-talk by the beacon can be neglected.
By semi-empirical study, we show that
the key rate of our scheme in \LEO satellite-to-ground \QKD is better than the pure \TDM and \WDM methods.
For zenith angle $\zeta \le 30^\circ$, the improvement are up at least $215\%$ for beam separation $L = 5$~m on a $400$~km altitude satellite using a response time $T_r$ equals $3$~dB of the whole system response bandwidth.
We also find that the lower the bandwidth, the higher the key rate improvement.
We also consider an alternative system that adds a pioneer beacon beam to existing pure \WDM systems.  We find that for zenith angle less than about $45^\circ$, this alternative setup performs \QKD at a rate of at least 90\% of our original proposal, making it an attractive engineering option in practice.
Further analysis of the system performance can be found in the Master thesis of one of the authors~\cite{chan_thesis}.
And we are going to report the effects of fine tuning the delay time $T_r$ and beam separation $L$ in our followup work.
Lastly, we stress that our method is also applicable to classical optical free-space communication as we do not any quantum property of the signal source.
 
\begin{acknowledgments}
 We would like to thank Hoi-Kwong Lo, Alan Pak Tao Lau, Chengqiu 
 Hu, Wenyuan Wang, and Gai Zhou for the discussion in optics. This work is supported by the RGC grant~17302019 of the Hong Kong SAR
 Government.
\end{acknowledgments}

\section*{Conflict of Interest}
 This paper is related to a patent application submitted by the authors.

\section*{Data Availability Statement}
 The data that support the findings of this study are available from the
 corresponding author upon reasonable request.

\appendix
\section{Wavefront Variance Calculation}
\label{Sec:sigma}

The calculation here is based on Tyson and Frazier's book~\cite{Tyson} and the paper by Devaney \emph{et al.}~\cite{Devaney:08}.
First, we consider the variance caused by the limitation of the bandwidth of the \AO system. 
It can be written as~\cite{Tyson}
\begin{equation}
    \sigma_\text{band}^2 = \int^\infty_0 |1-H(f,f_c)|^2 F(f)\,df,
    \label{E:App1_begin}
\end{equation}
where $f_c = 500$~Hz is the bandwidth of the \AO system, $f$ is a frequency variable,  
\begin{equation}
    H(f,f_c) = \left(1+\frac{if}{f_c}\right)^{-1}
\end{equation}
is the RC filter that used to model the system, and
\begin{equation}
 F(f) = 0.0326k^2 f^{-8/3}\int^{z(\zeta)}_0 v^{5/3}(z) C^2_n(z)\,dz
\end{equation}
is the power spectrum of the turbulence frequency.
Here $v(z)$ is the wind speed calculated using Eq.~\eqref{E:theta_0}.
Note that the bandwidth-limited wavefront variance is the same for all \AO systems studied in this paper.

Next, we discuss the chromatic effects that appears in \WDM systems.
The first contribution of chromatic aberration is due to diffraction.
The diffraction pattern of the beams at the receiving end depends on the wavelength~\cite{Devaney:08}.
When the beacon wavelength is $\lambda_b$, the variance on measuring the signal beam with wavelength $\lambda$ is~\cite{Devaney:08}
\begin{widetext}
\begin{equation}
        \sigma_\text{d}^2 = \frac{4.08}{\pi}k^2\int^L_0\int^\infty_0 K^{-8/3}
        \left\{1-\left(\frac{4}{KD}\right)^2\left[J_1\left(\frac{KD}{2}\right)\right]^2 \right\}
        \left[\cos\left(\frac{zK^2}{2k_b}\right)-\cos\left(\frac{zK^2}{2k}\right)\right]^2 C^2_n(z)\,dz \,dK,
\end{equation}
\end{widetext}
where $K$ is the spatial frequency and $k_b = 2\pi/\lambda_b$ is the wavenumber of the beacon.
The second chromatic contribution comes from path length error between the beams.
A \DM is only able to compensate error perfectly at a single wavelength.
This is because there is a path length difference between light beams with different wavelengths.
The corresponding wavefront variance is~\cite{Devaney:08}
\begin{equation}
    \sigma_\text{ch}^2 = 1.03\left(\frac{D}{r_0}\right)^{5/3} \epsilon^2(\lambda,\lambda_b) ,
\end{equation}
where
\begin{equation}
 \epsilon(\lambda,\lambda_b) = \frac{\lambda_b}{\lambda}\frac{n_s(\lambda) - n_s(\lambda_b)}{n_s(\lambda_b)-1}.
\end{equation}
In the above equation, $n_s$ is the refractive index which calculated at standard pressure and temperature base on the Ciddor's model~\cite{Ciddor:96}.

Lastly, we calculate the wavefront variance the caused by chromatic anisoplanatism.
Light waves with different wavelength travel different paths because of dispersion.
The isoplanatic error induced by this can be written as~\cite{Devaney:08}
\begin{equation}
    \sigma_\phi^2 = \left[\frac{\sin(\zeta)\Delta n}{\cos^2(\zeta)}\right]^{5/3} T_{5/3},
\end{equation}
where 
\begin{equation}
    \Delta n = |n_s(\lambda) - n_s(\lambda_b)|
\end{equation}
is the difference in refractive index and 
\begin{equation}
 T_{5/3} = 2.91k_b^2\sec(\zeta)\int^{h_\text{alt}}_0 I^{5/3}(h) C^2_n(h)\,dh,
\end{equation}
with $I(h)$ equals to the integral of the air density normalized to the value at sea level
\begin{equation}
   I(h) = \int^h_0 \alpha(z)\,dz.
\end{equation}
For simplicity, we only take integral of the troposphere in this paper as this layer contributes most to $I(h)$.
Specifically, we follow the web site of Shelquist~\cite{shelquist_2019} by using
the air density model
\begin{equation}
 h(\rho) = 44330.8 - 42266.5\rho^{0.234969}
 \label{E:App1_end}
\end{equation}
for $h \le 1.1\times 10^4$~m, where $\rho$ is measured in unit of kg/m$^3$.
Clearly, $h(\rho)$ is an invertible function.  By denoting its inverse function by $\rho(h)$, then $\alpha(h)$ is simply $\rho(h) / \rho(0)$.

\bibliographystyle{aipnum4-2}
\bibliography{qc78.5}
\end{document}